# Data on One-dimensional Vertical Free Swelling Potential of Soils and Related Soil Properties

Eyo E. U[1]*, Uche Onyekpe[2]

[1*]Faculty of Environment and Technology, Department of Geography and Environmental Management, Civil Engineering Cluster, University of the West of England, Bristol, United Kingdom.
[2]School of Science, Technology and Health, Department of Data Science, York St. John University, United Kingdom.
u.onyekpe@yorksj.ac.uk

Corresponding author's email address: *eyo.eyo@uwe.ac.uk*

**Abstract**

Most of the damaging geo-hazards recorded in modern history are caused by soil swelling or expansion. Therefore, proper evaluation of a soil's capacity to swell is very crucial for the achievement of a secure and safe foundation for civil infrastructure. Hence, in this brief, a dataset of standardised one-dimensional soil swelling on laboratory engineered and natural soils is reported. Corresponding properties of these soils such as moisture content, void ratio, specific gravity, unit weight, Atterberg limits, size fractions, soil activity index including the state of compaction of the soils where applicable, are also presented. The dataset is made up of a total of 395 data on soil swelling potential are reported. With regards to the corresponding soil properties, a total of 219 data records of soil specific gravity, 321 data of initial moisture content, 163 data of void ratio, 273 data of dry unit weight, 347 data of liquid limit, 347 data of plastic limit, 395 data of plasticity index, 209 data of activity index, 339 data of clay content, 174 data of silt content, 246 data of optimum moisture content, 228 data of maximum dry density and 347 data of Unified Soil Classification System (USCS). The procedures followed for the determination of soil swelling and related soil properties are mostly contained in American Standards for Testing of Materials (ASTM D 4546, ASTM D 698, ASTM D 1557 and ASTM D698), American Association of State Highways Transport Officials (AASHTO T 258 and AASHTO-T99) and Turkish Standards (TS1900). Finally, the dataset described herein are intended to aid geotechnical engineers and researchers who are involved in statistical correlation studies, data analytics, machine learning predictions using soft computing methods mostly aimed at evaluating soil expansion especially during the preliminary phases of soil investigation and foundation design.

## Introduction

Practicing ground engineering experts and stakeholders in geology and related disciplines have been seriously occupied with the challenges posed by swelling soils on foundations of structures and similar land development ventures. One of the reasons for this continuing interest is because in the past 50 decades, the reported cost of damage to infrastructure due to undesirable expansion has amounted to several billions of dollars. Moreover, soil swelling-related disaster has been regarded as one of the most devastating geo-hazards at least in modern history. Therefore, the proper identification and characterisation of an expansive soil's ability to swell is very vital for achieving a secure and safe construction ground for civil structures. In this article, dataset of one-dimensional swelling tests adopted to measure soil swelling on laboratory engineered and natural soils covering various regions on the globe are reported. The testing standards and procedures followed in the measurement of one-dimensional swelling are those enumerated in American Standards for Testing of Materials ASTM D 4546-14 [1], ASTM D 4546-96 [2], and American Association of State Highways Transport Officials AASHTO T 258 [3]. Slight modifications to the measurement procedures (such as the use of different surcharge loading and custom-made consolidation rings) reflecting special laboratory testing conditions and for the purposes of comparisons, are also reported. Related soil properties characterising the dataset includes moisture content, void ratio, specific gravity, unit weight, liquid limit, plastic limit, plasticity index, clay content, silt content, maximum dry unit weight, optimum moisture content and soil activity index all of which are known to bear either direct or indirect influences on soil. Determination of the state of compaction of the soils where applicable, were carried out based on the American Standards for Testing of Materials

ASTM D 698-07 [4], ASTM D 1557-07 [5], ASTM D698-00a [6], Turkish Standards (TS1900) [7], American Association of State Highways Transport Officials AASHTO-T99 [8] and a combination of both standard and modified efforts. A total of 395 data on soil swelling potential are reported. With regards to the corresponding soil properties, a total of 219 data records of soil specific gravity,  321 data of initial moisture content, 163 data of void ratio, 273 data of dry unit weight, 347 data of liquid limit, 347 data of plastic limit, 395 data of plasticity index, 209 data of activity index, 339 data of clay content, 174 data of silt content, 246 data of optimum moisture content, 228 data of maximum dry density and 347 data of Unified Soil Classification System (USCS) are presented. Finally, the dataset of one-dimensional soil swelling described herein are intended to aid geotechnical engineers and researchers who are involved in statistical corelation studies, data analytics, machine learning predictions using soft computing methods mostly aimed at evaluating soil expansion especially during the preliminary phases of soil investigation and foundation design.

**How the data was acquired**

Database of soils of low-to-high plasticity subjected to expansion under water were diligently and carefully compiled from literature. Soil swelling data were obtained through laboratory one-dimensional swelling test on soils using the consolidometer (or oedometer) according to standard measurement methods American Standards for Testing of Materials ASTM D 4546-14 [1], ASTM D 4546-96 [2], American Association of State Highways Transport Officials AASHTO T 258 [3] and slight modifications to these methods.

The data on the corresponding index properties of the soils namely moisture content, void ratio, specific gravity, unit weight, liquid limit, plastic limit, plasticity index, clay content, silt content, soil activity and compaction characteristics were also obtained using standard laboratory testing methods. The maximum dry unit weight and optimum moisture content were determined by following the American Standards for Testing of Materials ASTM D 698-07 [4], ASTM D 1557-07 [5], ASTM D698-00a [6], Turkish Standards (TS1900) [7], American Association of State Highways Transport Officials AASHTO-T99 [8] and a combination of both standard and modified efforts.

**Value of the data**

- This data is very useful to guide the geotechnical design of foundations of structures that are constructed on expansive soils
- Geotechnical engineers, researchers, and ground engineering experts who are involved with and interested in the accurate recognition and evaluation of expansive soil's capacity to swell during the preliminary stages of soil investigation, site characterisation and general soil expansion and ground movement research, will benefit from this data.
- This dataset can provide insight into the influence of index and intrinsic soil properties on swelling when adopting statistical correlation techniques, data analytics and predictions using soft computing methods.

**Description of data collection**

Both naturally occurring and laboratory-engineered soils are prepared and swelling test carried out to measure one-dimensional expansion. Undisturbed and disturbed soil materials are carefully inserted into oedometer rings and made to sit in the consolidometer between two porous stones with top and bottom of the porous stones lined with filter papers. Water is then gradually introduced into the oedometer (consolidometer) to inundate the soil samples and allowed to undergo free vertical swelling under a surcharge load for a minimum period of 24 hours until equilibrium is attained. On the other hand, standard laboratory testing procedures and indirect calculations are followed to obtain the index properties of the soils as follows:

- Liquid limits – using cone penetrometer (fall cone device) methods and a standard Casagrande cup.
- Plastic limit – the crumbling of moist soils through repeated remoulding into a small ball and manual rolling into 3mm thread.
- Plasticity index – obtained as a difference between liquid limit and plastic limit.
- Moisture content – oven drying method
- Void ratio – indirect determination by first calculating the volume of soil solids and then subtracting the volume of solids from the total volume to obtain volume of voids as a ratio
- Specific gravity of soils – using laboratory volumetric flask to determine the ratio of the weight of a given volume of soil at a specified temperature to the weight of an equal volume of water at the same temperature.
- Dry unit weight – indirect determination of the ratio of weight of dry soil to the total volume of the soil.

- Activity index – indirect determination by dividing plasticity index by the percentage of clay-sized fractions.
- Clay and silt content - standard testing methods for gradation (Particle-Size Distribution, PSD) of soils using sieve analysis.
- Optimum moisture content and maximum dry density – soil static and dynamic compaction test methods using standard laboratory moulds with application of standard and modified efforts.

The classification of soils is made according to the Unified Soil Classification System (USCS).

**Data description**

Dataset of 395 number of soils swelling measurements collected from one-dimensional vertical movement of soils and corresponding soil properties are presented. This dataset references studies spanning over 5 decades leading back to the period when important discourse on soil swelling-related disasters were first considered. The dataset consists of both naturally occurring (disturbed and undisturbed soils) and a mix of laboratory-engineered-soils. Based on the nature and format of reporting from the different sources, the soils are divided into groups with each having properties collated from the same experimental methods. The dataset for each group are saved in individual tabs within the .xlsx file with the tab names (notations) corresponding to the descriptions on Table 1. Table 2 summarizes the soil characteristics measured, indirectly obtained, and perceived to bear possible influences on the swelling behaviour of the soils when inundated with water and consolidated one-dimensionally. A total of 219 data records of soil specific gravity, 321 data of initial moisture content, 163 data of void ratio, 273 data of dry unit weight, 347 data of liquid limit, 347 data of plastic limit, 395 data of plasticity index, 209 data of activity index, 339 data of clay content, 174 data of silt content, 246 data of optimum moisture content, 228 data of maximum dry density, 347 data of Unified Soil Classification System (USCS) observations are contained in the dataset. The soil property attributes presented in Table 1, correspond to the features in the dataset presented at https://github.com/onyekpeu/Data-on-one-dimensional-vertical-free-swelling-potential-of-soils-and-related-soil-properties. More details on the data can be found on Table 3.

***Table 1** – Description of groups of datasets.*

| Dataset Notation | Soil description | Free swell test method | Surcharge (kPa) | No. of soil properties | Compaction state | Compaction Method |
|---|---|---|---|---|---|---|
| DS-1 | Low-medium plastic clayey soils | ASTM D 4546-14[1] | 1 | 13 | Compacted | Standard effort (ASTM D 698-07 [4] and modified effort (ASTM D 1557-07 [5] |
| DS-2 | Disturbed clay samples tested under different conditions | ASTM D 4546 [1] | 7 | 13 | Compacted | ASTM D698-07 [4] |
| DS-3 | Undisturbed clay samples from trial pits and boreholes | Fixed-ring oedometer | 7 | 7 | Non-compacted | NA |
| DS-4 | Laboratory engineered clay (mixture of bentonite and kaolinite in various proportions) | ASTM D 4546-96 [2] | 1 | 9 | Compacted | Turkish Standards (TS1900) [7] |
| DS-5 | Disturbed expansive clay soil samples | ASTM D4546-14 [1] | 1 | 9 | Compacted | ASTM D698-00a [6] |
| DS-6 | Disturbed expansive clay soil samples | Custom-made (oedometer cell, with a ring of | 0 | 9 | Compacted | Kolay and Singh [10] |
| | | 50 mm internal diameter and 30 mm height) | | | | |
| DS-7 | Disturbed expansive clay soil samples | standard fixed ring consolidometer using stainless steel rings, 75 mm inside diameter and 19 mm height | 7 | 10 | Compacted | Standard Proctor energy: 593.7 kJ/m3; Reduced Modified Proctor energy: 1616 kJ/m3; Modified Proctor energy: 2693.3 kJ/m3 [11] |
| DS-8 | Undisturbed expansive clay shales | ASTMD4546-96[2] | 7 | 7 | Non-compacted | NA |
| DS-9 | Disturbed expansive clay soils | Conventional oedometer | 6.9 | 6 | Compacted | Standard Proctor |
| DS-10 | Various disturbed soil samples | Custom-made (2 cm high consolidation ring | 6.9 | 5 | Compacted | AASHTO [8] |
| DS-11 | Laboratory engineered clay (Silica sand and artificial clays) | Custom-made | 6.9 | 7 | Compacted | AASHTO-T99 [8] |
| DS-12 | Laboratory engineered clay (mixture of bentonite and kaolinite in various proportions) | ASTM D4546-96[2] | 1.94 | 8 | Compacted | Statically to specified moisture content. |
| DS-13 | Undisturbed expansive soil samples | ASTM D4546 – 96 [2], Holtz and Gibbs (1956), [12] | 7 | 8 | Non-compacted | NA |

***Table 2.*** *Soil swelling percent and corresponding soil properties.*

| Soil property | Abbreviation/symbol | Unit |
|---|---|---|
| Swell percent | SP | % |
| Liquid limit | LL | % |
| Plastic limit | PL | % |
| Plasticity index | PI | % |
| Moisture content (initial) | Mc | % |
| Void ratio | e | - |
| Specific gravity | G | - |
| Dry unit weights | $\gamma_d$ | kN/m$^3$ |
| Activity index | A | - |
| Clay content | CC | % |
| Silt content | Sc | % |
| Optimum moisture content | OMC | % |
| Maximum dry density | MDD | $kN/m^3$ |
| Unified soil classification system | USCS | - |

***Table 3*** *– Further details on the dataset*

| | Information |
|---|---|
| Subject | Civil and Structural Engineering |
| Specific subject area | Soil swelling and properties of soils. |
| Type of data | Excel, xlsx |
| Data format | Raw, Filtered |
| Data source location | Global – Asia, Africa and Europe. |
| Data accessibility | Repository name: Github.com<br>Direct URL to data: https://github.com/onyekpeu/Data-on-one-dimensional-vertical-free-swelling-potential-of-soils-and-related-soil-properties- |

**Experimental design and methods**

Rigorous systematic literature studies were conducted to obtain the dataset. The literature search included data spanning across research articles and reports with experimental investigations conducted on various soil samples. The methodology of data collection involved thorough screening of articles' titles, keywords and abstracts that met the criteria of one-dimensional swelling for both compacted and non-compacted soils. Most of the collected data were derived from article tables and where data were only presented on graphs, WebPlotDigitizer [13] was used for data extraction.

Relevant data of soil properties not included in the original data such as the Plasticity Index ($PI$) and Plastic Limits ($PL$) of the soil were derived from the parameters that were given as follows:

$$PI = LL - PL \tag{1}$$

Or

$$PL = LL - PI \tag{2}$$

where $LL$ refers to the liquid limit.

In areas where soil Activity index, ($A$) was not given, this was calculated as follows:

$$A = \frac{PI}{2CF} \tag{3}$$

where $CF$ is the clay-sized fraction

To allow for an easy and extensive assessment and analysis of the degree of swelling of the different soils and their properties obtained from various sources, entries into each column of the dataset were manually synchronised to account for missing data through the removal of such affected rows and columns.

**CRediT author statement**

***Eyo Eyo U:*** *Conceptualization, Methodology, Software, Data curation, Writing- Original draft preparation.* ***Uche Onyekpe****: Visualization, Investigation, Reviewing and Editing.*

**Declaration of interests**

The authors declare that they have no known competing financial interests or personal relationships that could have appeared to influence the work reported in this paper.